\documentclass[%
 aip,
 amsmath,amssymb,
 reprint,%
]{revtex4-1}

\usepackage{graphicx}
\usepackage{dcolumn}
\usepackage{bm}

\usepackage[utf8]{inputenc}
\usepackage[T1]{fontenc}
\usepackage{mathptmx}
\usepackage{etoolbox}

\makeatletter
\def\@email#1#2{%
 \endgroup
 \patchcmd{\titleblock@produce}
  {\frontmatter@RRAPformat}
  {\frontmatter@RRAPformat{\produce@RRAP{*#1\href{mailto:#2}{#2}}}\frontmatter@RRAPformat}
  {}{}
}%
\makeatother
\usepackage{subfigure}
\usepackage{mathrsfs}
\usepackage[T1]{fontenc}
\usepackage{graphicx}
\usepackage{dcolumn}
\usepackage{bm}
\usepackage{color}
\usepackage{times}
\usepackage[colorlinks=true, citecolor=blue, linkcolor=magenta, anchorcolor=black, urlcolor=blue]{hyperref}
\usepackage{siunitx}
\usepackage{overpic}
\usepackage{braket}
\usepackage{changes}
\usepackage{float}

\newcommand{\addrA}{Beijing Key Laboratory of Fault-Tolerant Quantum Computing,\\ Beijing Academy of Quantum Information Sciences, Beijing 100193, China} 
\newcommand{\addrB}{Beijing National Laboratory for Condensed Matter Physics,\\ Institute of Physics, Chinese Academy of Sciences, Beijing 100190, China} 
\newcommand{\addrC}{School of Physical Sciences, University of Chinese Academy of Sciences, Beijing 100049, China} 
\newcommand{\addrD}{Hefei National Laboratory, Hefei 230088, China} 
\newcommand{\addrE}{Songshan Lake Materials Laboratory, Dongguan 523808, Guangdong, China} 

\begin{document}
\title{Microwave Engineering of Tunable Spin Interactions with Superconducting Qubits}

\author{Kui Zhao}
    \thanks{These authors contributed equally to this work.}
	\affiliation{\addrA}

\author{Ziting Wang}
    \thanks{These authors contributed equally to this work.}
	\affiliation{\addrA}
	
\author{Yu Liu}
    \email{yuliu@iphy.ac.cn}
    \thanks{These authors contributed equally to this work.}
	\affiliation{\addrB}
	\affiliation{\addrC}
	
\author{Gui-Han Liang}
    \thanks{These authors contributed equally to this work.}
	\affiliation{\addrB}
	\affiliation{\addrC}
	
\author{Cai-Ping Fang}
	\affiliation{\addrB}
	\affiliation{\addrC}
	
\author{Yun-Hao Shi}
	\affiliation{\addrB}
	
\author{Lv Zhang}
	\affiliation{\addrB}
	\affiliation{\addrC}   
    
\author{Jia-Chi Zhang}
	\affiliation{\addrB}
	\affiliation{\addrC}
    
\author{Tian-Ming Li}
	\affiliation{\addrB}
	\affiliation{\addrC}	

\author{Hao Li}
	\affiliation{\addrA}

\author{Yueshan Xu}
	\affiliation{\addrA}
	
\author{Wei-Guo Ma}
	\affiliation{\addrB}
	\affiliation{\addrC}

\author{Hao-Tian Liu}
	\affiliation{\addrB}
	\affiliation{\addrC}

\author{Jia-Cheng Song}
	\affiliation{\addrB}
	\affiliation{\addrC}

\author{Zhen-Ting Bao}
	\affiliation{\addrB}
	\affiliation{\addrC}

\author{Yong-Xi Xiao}
	\affiliation{\addrB}
	\affiliation{\addrC}

\author{Bing-Jie Chen}
	\affiliation{\addrB}
	\affiliation{\addrC}

\author{Cheng-Lin Deng}
	\affiliation{\addrB}
	\affiliation{\addrC}
	
\author{Zheng-He Liu}
	\affiliation{\addrB}
	\affiliation{\addrC}	

\author{Yang He}
	\affiliation{\addrB}
	\affiliation{\addrC}
	
\author{Si-Yun Zhou}
	\affiliation{\addrB}
	\affiliation{\addrC}	
	
\author{Xiaohui Song}
	\affiliation{\addrB}
	\affiliation{\addrC}

\author{Zhongcheng Xiang}
	\affiliation{\addrB}
	\affiliation{\addrC}
	\affiliation{\addrD}

\author{Dongning Zheng}
	\affiliation{\addrB}
	\affiliation{\addrC}
	\affiliation{\addrD}
	\affiliation{\addrE}
	
\author{Kaixuan Huang}
    \email{huangkx@baqis.ac.cn}
	\affiliation{\addrA}
	
\author{Kai Xu}
	\affiliation{\addrA}
	\affiliation{\addrB}
	\affiliation{\addrC}
	\affiliation{\addrD}
	\affiliation{\addrE}
    
\author{Heng Fan}
    \email{hfan@iphy.ac.cn}
	\affiliation{\addrA}
	\affiliation{\addrB}
	\affiliation{\addrC}
	\affiliation{\addrD}
	\affiliation{\addrE}

\begin{abstract}
Quantum simulation has emerged as a powerful framework for investigating complex many-body phenomena. A key requirement for emulating these dynamics is the realization of fully controllable quantum systems enabling various spin interactions. Yet, quantum simulators remain constrained in the types of attainable interactions. Here we demonstrate experimental realization of multiple microwave-engineered spin interactions in superconducting quantum circuits. By precisely controlling the native XY interaction and microwave drives, we achieve tunable spin Hamiltonians including: (i) XYZ spin models with continuously adjustable parameters, (ii) transverse-field Ising models, and (iii)  Dzyaloshinskii-Moriya interacting systems. Our work expands the toolbox for analogue-digital quantum simulation, paving the way for exploring a wide range of exotic quantum spin models.
\end{abstract}
\maketitle


Interacting quantum spin systems can host intriguing many-body dynamical phenomena, spanning from thermalization dynamics~\cite{Andersen2025} to topological states~\cite{Kiczynski2022} and non-equilibrium phases~\cite{Zhang2022}. Quantum simulation, employing highly controllable quantum systems to mimic the behaviors of target quantum systems, has emerged as a powerful tool for investigating these interesting phenomena~\cite{Georgescu2014}. At the core of quantum simulation lies the ability to engineer desired spin interactions. For instance, XY interactions with controllable Peierls phase can be used to synthesize artificial gauge fields to emulate the fractional quantum Hall effect~\cite{Wang2024},  magnetic vector potentials~\cite{Rosen2024} and topological phase transition~\cite{Jotzu2014}, significantly expanding the range of attainable Hamiltonians for studying many-body physics.

Leading quantum simulators include trapped ions~\cite{Monroe2021,Kranzl2023,Qiao2024}, ultracold atoms~\cite{Gross2017}, photonic systems~\cite{AspuruGuzik2012}, and superconducting quantum circuits~\cite{Houck2012}. These experimental platforms feature unique capabilities for engineering specific Hamiltonian through tunable system parameters. However, the analogue nature of current quantum simulators imposes fundamental limits on their experimental versatility, due to fixed native interactions. While digital quantum simulators~\cite{Heras2014,Salathe2015,Fauseweh2024} promise greater flexibility through universal gate sets, their application to Hamiltonian simulation remains severely limited by the demanding circuit depths and the accumulation of gate errors in near-term devices.
An alternative approach to simulate interactions in spin models is to apply periodic drives~\cite{Sameti2019, Rosenberg2024, Nguyen2024, Liu2025}, significantly expanding the class of simulatable quantum systems.

Among various physical implementations, superconducting quantum circuits have emerged as a promising platform for quantum simulation, leveraging recent progress in coherence times, scalable architectures, and precise control~\cite{Koch2007, Yan2018, Krantz2019, Liang2023}. Specifically, transmon qubits offer exceptional tunability: their transition frequencies can be controlled via flux bias, while the XY interactions between neighboring qubits can be precisely adjusted through tunable couplers. In this work, we demonstrate programmable engineering of diverse spin interactions using only single-qubit gates and native XY interactions of the system in superconducting qubits. We develop an efficient and compact experimental framework to characterize and calibrate the phase correlations of transverse microwave fields, both individually and collectively. By precisely controlling the temporal interval between single-qubit gates, we realize exact dynamics governed by anisotropic Heisenberg interactions and transverse Ising interactions in a 2-qubit system, respectively. Furthermore, we synthesize and experimentally validate the Hamiltonian featuring tunable ratio between XY and Dzyaloshinskii-Moriya (DM) interactions~\cite{Dzyaloshinsky1958, Moriya1960} in an 8-qubit system.

\begin{figure}[!htbp]
	\includegraphics[width = 0.40\paperwidth]{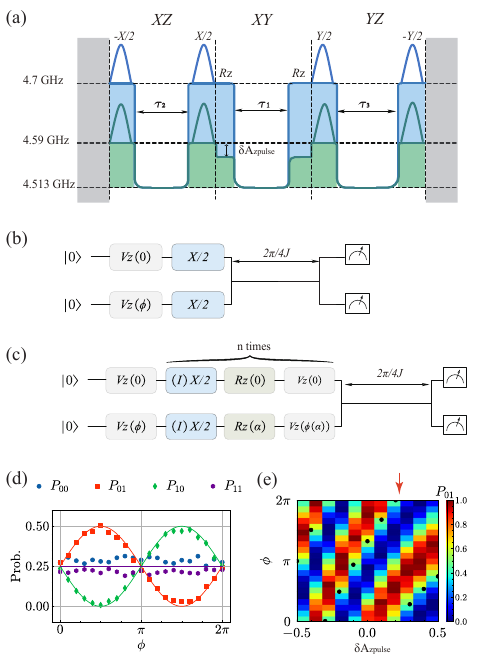}
	\caption{Implementation of the XYZ spin interaction. (a) Pulse sequence applied to two qubits for realizing the XYZ spin interaction. 
	Gaussian-shaped microwave pulses implement the single-qubit rotations $R^{x}(\pi/2)$ and $R^{y}(\pi/2)$, corresponding to X/2 and Y/2 gates, respectively, with a gate length of 20 ns. Z pulses, indicated by blue (green) solid lines for individual qubits, switch the qubits between their idle and resonant frequencies. At the idle frequency, these pulses enable single-qubit gate operations, while at the resonant (work) frequency, they bring the qubits into resonance to facilitate XY interactions. The blue (green) shaded regions associated with the Z pulses represent the phase accumulation resulting from the reference frame transition between idle and working frames.
	This accumulated phase must be an integer multiple of $2\pi$ for each qubit and is thus compensated in experiments using dedicated physical Z (Rz) gates.
	(b) Protocol for calibrating the initial qubit phase difference.
	Here, $V_{\mathrm{z}}$ denotes a virtual Z gate implemented via software-controlled adjustment of the microwave phase reference frame. The combined action of the $V_{\mathrm{z}}$ gate and subsequent X/2 gates prepares the initial state as the product state,  $\frac{1}{\sqrt{2}}(|0\rangle + |1\rangle) \otimes \frac{1}{\sqrt{2}}(|0\rangle + e^{i\phi}|1\rangle)$.
	(c) Protocol for calibrating the physical Z ($R_{\mathrm{z}}$) gate amplitude. The cycle is repeated for n times before we measure the qubits in the Z basis. 
	(d) Experimental results for calibrating qubit phase difference. (e) Experimental results for calibrating physical Z gate via a Z-pulse amplitude offset $\delta  A_{\mathrm{zpulse}}$.}
    \label{fig:XYZPulse}
\end{figure}

We begin by considering a general superconducting circuit system comprising $N$ transmon qubits with tunable couplers between all neighboring qubit pairs, under rotating wave approximation (RWA) the effective Hamiltonian can be written as ($\hbar=1$)~\cite{Yan2018}:
\begin{equation}
    \hat{H}=\sum_{i}\frac{1}{2}\omega_j\sigma_j^z+\sum_{\langle i,j\rangle}\hat{H}_{i,j}^{XY},
\end{equation}
with
\begin{equation}
    \hat{H}_{i,j}^{XY}=\frac{1}{2}J_{i,j}\left(\sigma_i^x\sigma_j^x+\sigma_i^y\sigma_j^y\right)=J_{i,j}\left(\sigma_i^+\sigma_j^-+\sigma_i^-\sigma_j^+\right),
\end{equation}
where $\sigma_j^{x,y,z}$ are Pauli matrices for the $j$-th qubit, $\sigma_j^{\pm} = (\sigma_j^x \pm \mathrm{i}\sigma_j^y)/2$ denote the lower and rising operators, $\omega_j$ is the qubit frequency, and $J_{i,j}$ is the XY coupling strength between nearest-neighbor qubit pairs. 

Building on techniques from nuclear magnetic resonance~\cite{Vandersypen2005}, solid-state spin systems~\cite{Ryan2010}, and Rydberg atom arrays~\cite{Geier2021, Scholl2022, Nishad2023, kuji2024proposalrealizingquantumspinsystems}, we can generate various spin interactions through global frame rotations. Specifically, applying transverse $\pi/2$ rotations yields the XZ interaction~\cite{Heras2014},
\begin{equation}
    \hat{H}_{i,j}^{XZ}=R^x_{i,j}(\pi/2)\hat{H}_{i,j}^{XY}R^x_{i,j}(-\pi/2)=\frac{1}{2}J_{i,j}\left(\sigma_i^x\sigma_j^x+\sigma_i^z\sigma_j^z\right),
\end{equation}
and the YZ interaction
\begin{equation}
    \hat{H}_{i,j}^{YZ}=R^y_{i,j}(\pi/2)\hat{H}_{i,j}^{XY}R^y_{i,j}(-\pi/2)=\frac{1}{2}J_{i,j}\left(\sigma_i^y\sigma_j^y+\sigma_i^z\sigma_j^z\right),
\end{equation}
where $R^{\mathbf{n}}_{i,j}(\theta)=R^{\mathbf{n}}_{i}(\theta)R^{\mathbf{n}}_{j}(\theta)$ denotes the global rotating operator along the $\mathbf{n}-$axis. Moreover, the local longitudinal rotation enables the phase difference between spin operators~\cite{Nishad2023}:
\begin{equation}\label{eq:XYDM}
    \begin{aligned}
        \hat{H}_{i,j}^{XY}(\Delta\phi_{ij})
        &= R_{i,j}^z(\phi_i,\phi_j)\hat{H}_{i,j}^{XY}R_{i,j}^z(-\phi_i,-\phi_j)\\
        &= \frac{1}{2}J_{i,j} \left(e^{\mathrm{i}\Delta \phi_{ij}} \sigma_i^{+} \sigma_j^{-}+e^{-\mathrm{i}\Delta \phi_{ij}} \sigma_i^{-} \sigma_j^{+}\right)\\
        &=  \frac{1}{2}J_{i,j} \left(\cos (\Delta \phi_{ij})(\sigma _{i}^x \sigma _{j}^x + \sigma _{i}^y \sigma _{j}^y) 
        \right. \\
         & \qquad \qquad \left. + \sin (\Delta \phi_{ij})(\sigma _{i}^x \sigma _{j}^y - \sigma _{i}^y \sigma _{j}^x)  \right),
    \end{aligned}
\end{equation}
with $R_{i,j}^z(\phi_i,\phi_j)=R_i^z(\phi_i) R_j^z(\phi_j)$, and  phase difference $\Delta\phi_{ij}=\phi_j-\phi_i$. Experimentally, the local rotations can be effectively achieved by virtual Z ($V_{\mathrm{z}}$)  gates~\cite{McKay2017}. 
When periodic gate sequences with total period $T_c$ are applied to the system, fragmenting the Hamiltonian into multiple components $\{\hat{H}_i\}$, the system dynamics in the high-frequency regime ($J_{i,j} T_c\ll 2\pi$) is governed by an averaged Hamiltonian.
To the lowest order, the effective Hamiltonian takes the form~\cite{Nishad2023}:
\begin{equation}
\hat{H}_{\textrm{eff}} = \frac{1}{T_c}\sum_i \tau_i \hat{H}_i,
\end{equation}
where $\tau_i$ represents the time allocation for each Hamiltonian component.By combining with various interactions under microwave control, this framework enables the engineering of spin interaction forms that go beyond the native XY interactions.


\begin{figure}[!htbp]
	\includegraphics[width = 0.4\paperwidth]{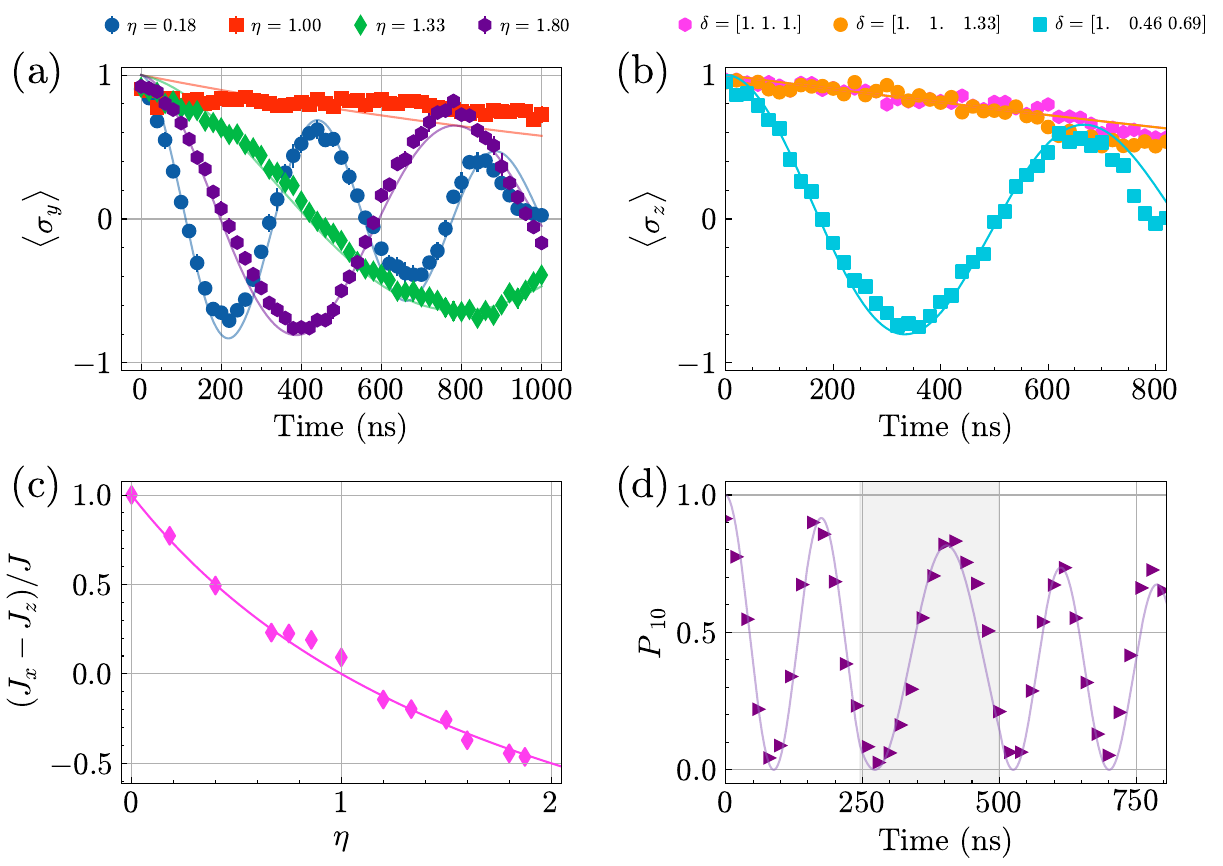}
	\caption{Experimental results of the XYZ interaction in a two-qubit system. (a) Time evolution of $y$-magnetization for the $\ket{++}_y$ initialized state under variable total interaction time ($\tau_1+\tau_2+\tau_3$) with fixed $\tau_1/\tau_2$ ($J_x=J_y$), demonstrating dynamics governed by the XXZ interaction at anisotropy parameters $\eta=J_z/J_x$, $\eta = \{0.18, 1.0, 1.33, 1.8\}$. (b) The z-magnetization dynamics initialized in a $z$-polarized state, contrasting XXZ (\(J_x = J_y \neq J_z\), partial $U(1)$ symmetry) and XYZ (\(J_x \neq J_y \neq J_z\), full symmetry breaking) regimes. We use $\delta=[1, J_y/J_x, J_z/J_x]$ to label the parameters.
    (c) Normalized oscillation frequency $J_x-J_z/J_0$ of $y$-magnetization in XXZ model versus anisotropy parameter \(\eta\), exhibiting quantitative agreement between experimental data (circles) and theoretical predictions (solid line).  (d) Time-dependent population $P_{10}$ under alternating Hamiltonian control: Gray-shaded regions implement \(H_{XXZ}\) via pulse modulation (\(\tau_1 = \tau_2 = \tau_3\)), while unshaded intervals restore native \(H_{XY}\) dynamics. The data points show experimental results,
    while the solid curves represent the numerical results incorporating qubit decoherence effects ($T_1\sim 10$~$\mu$s and $T_2^*\sim1.5$~$\mu$s, obtained at the resonant operating frequency.). The error bars represent one standard deviation.}
    \label{fig:XYZ}
\end{figure}

First, we demonstrate the realization of the XYZ spin interaction in a two-qubit system described by the Hamiltonian:
\begin{equation}
\hat{H}_{i,j}^{XYZ}=  J_x \sigma_i^x \sigma_j^x+ J_y \sigma_i^y \sigma_j^y + J_z \sigma_i^z \sigma_j^z,
\end{equation}
where $J_{x}$, $J_y$ and $J_z$ denote the coupling of spins along the $x$, $y$ and $z$ direction, respectively. To engineer the interaction, we employ a global gate sequence $\{-X/2,\, X/2,\, Y/2,\, -Y/2\}$, interleaved with resonant evolution periods $\tau_2$, $\tau_1$ and $\tau_3$, with $T_c=\sum_i\tau_i$, which respectively generate the XZ, XY, and YZ interaction components.  Importantly, these components commute in the two-spin case, enabling exact realization of the target dynamics within a single control period. The effective parameters are 
$J_x=J_{i,j}(\tau_1+\tau_2)/2T_c$, $J_y=J_{i,j}(\tau_2+\tau_3)/2T_c$, and $J_z=J_{i,j}(\tau_1+\tau_3)/2T_c$.


The pulse sequence used to implement the XYZ spin interaction on qubits $Q_1$ (initialized at 4.7~GHz) and $Q_2$ (initialized at 4.59~GHz) is illustrated in Fig.~\ref{fig:XYZPulse}(a).
(i) To achieve the global control $(-X/2)\otimes(-X/2)$, we need carefully calibrate the phase difference between the qubits. The initial state is prepared using $R^x(\pi/2)\otimes R^{\mathbf{n}}(\pi/2)$, where $\mathbf{n}$ can be adjusted by $V_\mathrm{z}$ gate~\cite{Song2017} , and the qubits then undergo resonant evolution for a duration $t=\pi/4|J|$ (Fig.~\ref{fig:XYZPulse}(b)) under varying virtual Z gate phase $\phi$. For $J/2\pi=-3$~MHz in our experiments, the population of $\ket{01}$ follows a sinusoidal dependence $(1+\sin(\phi))/4$, when $\mathbf{n}$ is aligned along the 
x-axis. As plotted in Fig.~\ref{fig:XYZPulse}(d), we show the experimental results when the microwave phases are aligned. (ii) Subsequently, the qubits are biased to 4.513~GHz for a resonant evolution with a duration of $\tau_2$, after which they are returned to their idle frequencies for a global $X/2\otimes X/2$ rotation. Dynamical phase corrections are applied to ensure accurate global control~\cite{Xiang2023,Huang2021,sciadv.aba4935KaiXu}. Prior to the next resonance, a phase-compensation padding is applied to $Q_1$ to compensate for the relative frame phase difference with respect to the resonant frequency, while a physical Z gate (labeled as $R_\mathrm{z}$) of equal duration is simultaneously applied to $Q_2$. The mechanism underlying phase compensation is elaborated upon in the supplementary material.
The $R_\mathrm{z}$ gate is implemented by detuning the qubit from its idle frequency via a Z-pulse amplitude offset $\delta  A_{\mathrm{zpulse}}$, which induces a frequency shift $\delta\omega = \alpha \delta  A_{\mathrm{zpulse}}$ (where $\alpha$ is the qubit's flux sensitivity), thereby accumulating a controlled phase shift. The $R_\mathrm{z}$ amplitude, i.e., $\delta  A_{\mathrm{zpulse}}$, is precisely calibrated via a post-resonance evolution of duration $\pi/4|J|$, as depicted in Fig.~\ref{fig:XYZPulse}(c). 
This protocol enables comprehensive calibration across multiple cycles, simultaneously ensuring accurate $R_\mathrm{z}$ gate implementation while correcting dynamical phase shifts caused by deviations from the qubit's idling frequency during $R_\mathrm{z}$ operations.
Microwave phase alignment is subsequently validated, with representative results provided in Fig.~\ref{fig:XYZPulse}(e). (iii) The padding and physical Z gate with the same duration are applied to qubits to align the microwave phase before the third resonance. (iv) Finally, the qubits are biased on resonance for $\tau_3$, and are returned to idle points followed by a rotation of $Y/2\otimes Y/2$.

\begin{figure}[htbp]
	\includegraphics[width = 0.30\paperwidth]{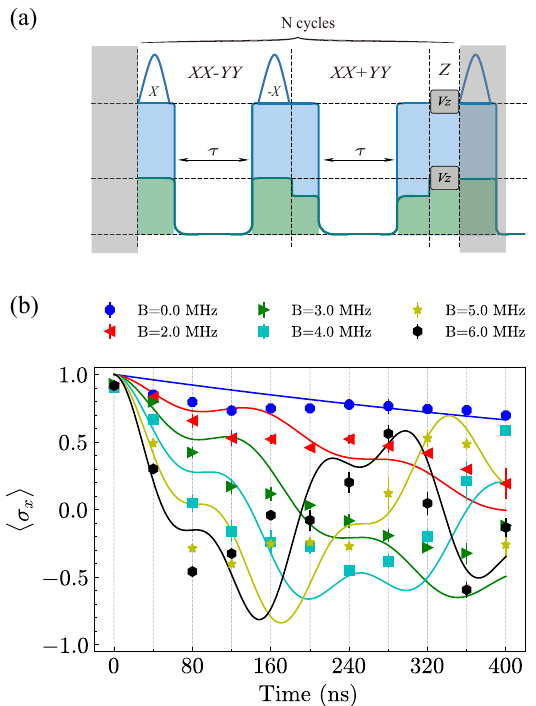}
	\caption{Experimental realization of the transverse-field Ising interaction in a two-qubit system. (a) Pulse sequence applied to two qubits for realizing the transverse-field Ising interaction, the coupling strength is tuned to be \(J/2\pi =-3 \) MHz. (b) Time evolution of the $x$-magnetization initialized in $\ket{++}_x$ under various transverse fields. Experimental data (circles), measured over successive 40-ns sequence cycles, exhibit qualitative agreement with numerical simulations (solid curves) incorporating qubit decoherence. Error bars denote the standard error of the mean across three experimental repetitions.}
    \label{fig:XXIsing}
\end{figure}

\begin{figure*}[t]
    \centering
    \includegraphics[width=0.75\linewidth]{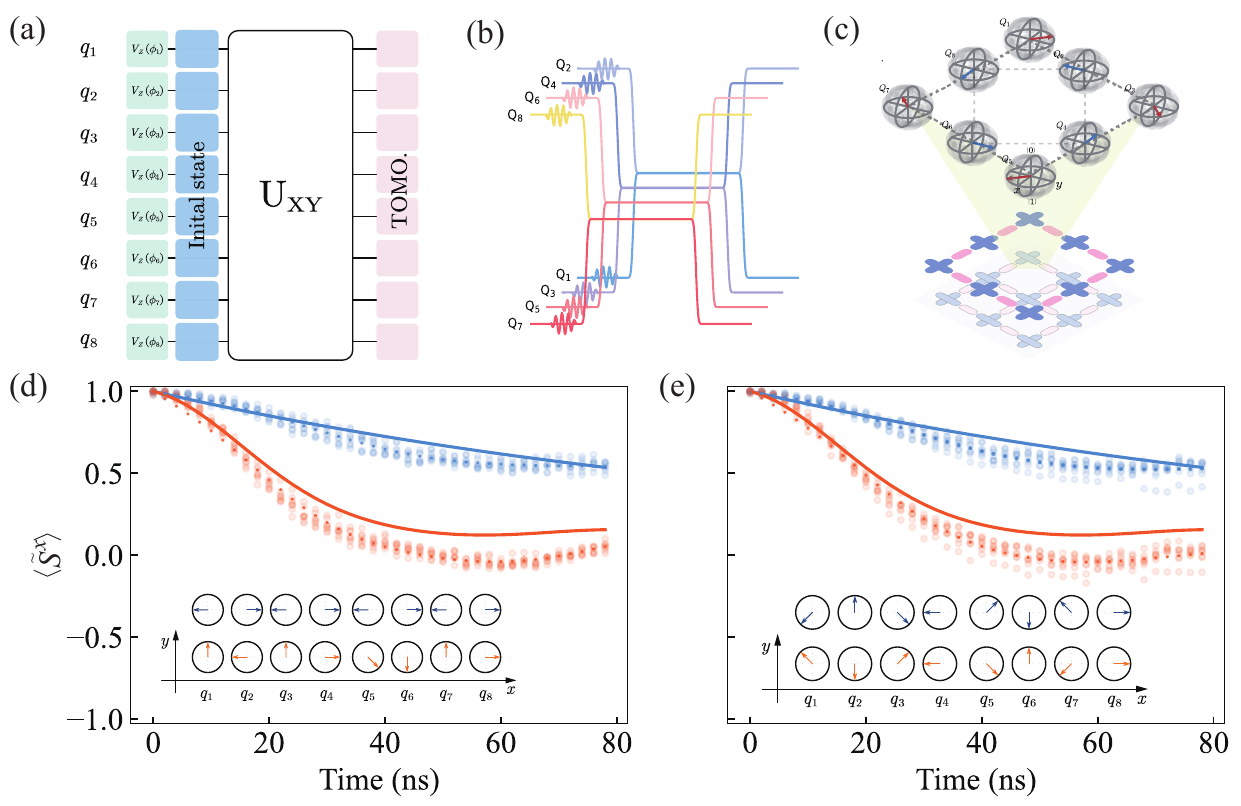}
    \caption{Implementation of the Dzyaloshinskii-Moriya (DM) spin interaction. (a) Protocol for realizing the tunable XY and DM interactions in the effective Hamiltonian. An 8-qubit ring with periodic boundary conditions is constructed, and virtual Z gates are employed to emulate \( R_z(\phi) \) rotations. These rotations encode tunable spin couplings, where virtual phase differences between adjacent qubits are set to \( \pi/2 \) or \( \pi/4 \), directly controlling the relative XY+DM interaction strength (\( G/D = 0 \) and \( G/D = 1 \), respectively).  
     (b) Example pulse sequence for the 8-qubit system, comprising state preparation, resonant evolution, and tomography-based measurement. (c) Experimental setup on a superconducting quantum processor with a 3×3 transmon qubit array. The central qubit is detuned to a low frequency relative to the other eight qubits to isolate an 8-qubit ring configuration. During resonance, inter-qubit coupling strengths are tuned to -8 MHz via tunable couplers. (d) Time-dependent local magnetization $\langle \tilde{S}^x(t) \rangle$ for $G/D = 0$ (pure 
    DM interaction). Experimental results (points) for the eight-qubit ring show the time evolution of the local magnetization $\langle \tilde{S}^x(t) \rangle$ when initialized in the zero-energy eigenstate $|\psi_0\rangle$. The inset depicts two distinct initial spin configurations in the $x$-$y$ plane: the eigenstate (blue) and a non-eigenstate (yellow). The eigenstate exhibits markedly slower decay dynamics compared to the non-eigenstate. (e) Similar to (d) but for \( G/D = 1 \), the non-eigenstate configuration (yellow), initialized with a mirror - symmetric spiral spin texture as opposed to the eigenstate’s spiral spin texture (blue), exhibits significantly faster relaxation dynamics.  
    The solid curves in panels (d) and (e) represent the numerical results accounting for qubit decoherence parameters ($T_1 \sim 10~\mu$s, $T_2^* \sim 1.5~\mu$s, mean value of parameters at the resonant operating frequency.).
    }    \label{fig:DMz}
\end{figure*}


To characterize the engineered Hamiltonian, we experimentally track the two-qubit magnetization dynamics, which are governed by Hamiltonians with distinct symmetries. 
By implementing a gate sequence with variable total interaction time, at fixed anisotropy ratio $\tau_1/\tau_2$, we realize XXZ dynamics where $J_x=J_y$, with $\eta=J_z/J_x$ serving as a tunable control parameter. 
As shown in Fig.~\ref{fig:XYZ}(a), initialized as $\ket{++}_y=R^y(\pi/2)\otimes R^y(\pi/2)\ket{00}$, the XXX interaction case ($\eta=1$) possesses SU(2) symmetry, which leads to frozen magnetization dynamics (red curve) due to spin rotational symmetry conservation. When the symmetry is broken by tuning $\eta\neq 1$, the system realizes an XXZ Hamiltonian that exhibits $y$-magnetization dissipation while maintaining $U(1)$ symmetry about the $z$-axis. This residual $U(1)$ symmetry persists until full anisotropy is introduced through the condition $J_x\neq J_y\neq J_z$, which completely breaks rotational invariance. As shown in Fig.~\ref{fig:XYZ}(b), the engineered XYZ Hamiltonian induces clear z-magnetization oscillation (cyan curve), demonstrating the absence of any conserved spin components in the fully anisotropic case. Moreover, we demonstrate continuous control of $\eta$ across $0 < \eta < 2$ by varying total interaction time while maintaining fixed $\tau_1/\tau_2$ ratios. Figure~\ref{fig:XYZ}(c) shows excellent agreement between measured $\eta$-dependence and theoretical prediction $J_x(\eta)-J_z(\eta) = 2J(1-\eta)/(2+\eta)$, confirming precise Hamiltonian engineering.  To further demonstrate dynamic control, we initialize the system in $\ket{10}$ and monitor the time-dependent population $P_{10}$. Under the native XY Hamiltonian  $H_{\mathrm{XY}}$, coherent oscillations between $\ket{10}$ and $\ket{01}$ occur at coupling strength $J$ (Fig.~\ref{fig:XYZ}(d)). When an additional isotropic XXX interaction is introduced during the time interval from 0.25 to 0.5~$\mu$s, we observe a reduction in the oscillation frequency by a factor of 0.65(2). This measured value agrees well with the theoretically predicted scaling factor of $2/3$.

Next, we focus on the realization of the transverse-field Ising interaction
\begin{equation}
    \hat{H}_{i,j}^{TI}=J_{t}\sigma_i^x\sigma_j^x+\frac{B}{2}\sum_i\sigma_i^z,
\end{equation}
where $J_{t}$ controls the spin-exchange strength with $J_t/2\pi=-3$~MHz and $B$ represents an effective transverse field. As shown in Fig.~\ref{fig:XXIsing}(a), the XX interaction component is realized through two sequential applications of the native XY interaction (each lasting $\tau=T_c/2$), with a carefully timed $\pi$-pulse applied to qubit $Q_i$ between the interaction intervals. This staggered sequence provides a cancellation of unwanted YY terms while constructively enhancing the desired XX coupling over a complete cycle. The transverse field term is implemented via virtual Z rotations with $\phi=BT_c$ enabled by dynamic frame tracking. Similar to the XYZ interaction case, padding and physical Z gates are used to compensate the relative microwave phase when switching between idle points and work points. In Fig.~\ref{fig:XXIsing}(b), we show the experimental results of the dynamics governed by the transverse-field Ising interaction over 10 cycles. 
Note that the observed deviations in  Fig.~\ref{fig:XXIsing}(b) primarily stem from residual $R_\mathrm{z}$ gate compensation inaccuracies, minor dynamic phase accumulation errors, and coherence limitations scaling with cycle count.

Here and below, we focus on the DM interaction~\cite{Dzyaloshinsky1958, Moriya1960}, which is characterized by antisymmetric interaction term $\sum_{i\neq j}\vec{D}_{ij}\cdot(\vec{\sigma}_i\times \vec{\sigma}_j)$, with the interaction strength vector $\vec{D}_{ij}$. In contrast to the Heisenberg exchange which energetically favors collinear spin alignment, the DM interaction represents a chiral exchange mechanism that promotes non-collinear spin configurations. This anisotropic interaction requires broken inversion symmetry and emerges from spin-orbit coupling in localized spin systems~\cite{Kuepferling2023}. 
While parametric modulation techniques have successfully engineered DM interactions in superconducting systems~\cite{Roushan2017,Wang2019,LiuWuxinAPL2020}, these approaches lack flexibility for realizing diverse spin interactions. We here employ a hybrid digital-analog method combining single-qubit gates with resonant XY evolution to synthesize programmable XY+DM interactions.
We utilize a 9-qubit superconducting processor with 12 tunable couplers~\cite{Wang2023Escaping,Wang2024Quantum,Wang2024Demonstration} to synthesize the effective
Hamiltonian combining isotropic XY interaction and DM interaction of the form~\cite{qute.201900121,PhysRevB.78.214414,PhysRevA.89.042306,Jafari2011},  
\begin{equation}\label{Eq_DM_XY}
    \hat{H}_{\textrm{XY+DM}} = \sum_{\langle i,j\rangle}\left[G(\sigma_i^x\sigma_j^x + \sigma_i^y\sigma_j^y) + D(\sigma_i^x\sigma_j^y - \sigma_i^y\sigma_j^x)\right],
\end{equation}
where $G$ and $D$ respectively denote XY and DM interaction strengths. This Hamiltonian is synthesized through local modulation $\sum_{\langle ij \rangle} R_{i,j}^z(\phi_i,\phi_j)\hat{H}_{i,j}^{XY}R_{i,j}^z(-\phi_i,-\phi_j)$ (Eq.~(\ref{eq:XYDM})) that transforms the native XY interaction into:
\begin{equation}\label{eq:H005}
\begin{aligned}
H_{\text{XY+DM}} =& \sum_{\langle ij \rangle} J_{ij} \left(\cos\Delta\phi_{ij}(\sigma_i^x\sigma_j^x + \sigma_i^y\sigma_j^y) + \right. \\&
\left. \sin\Delta\phi_{ij}(\sigma_i^x\sigma_j^y - \sigma_i^y\sigma_j^x)\right), 
\end{aligned}
\end{equation}  
with $\Delta\phi_{ij} \equiv \phi_j - \phi_i$ representing the relative phase of $R_z^{(i)}(\phi_i) = \exp(-i\sigma_i^z\phi_i/2)$ rotations. The Hamiltonian with tunable ratio $G/D$ is realized through gate sequences employing local $R_\mathrm{z}$ modulations~\cite{Nishad2023,kuji2024proposalrealizingquantumspinsystems}. By fixing the phase difference $\Delta\phi_{ij} \equiv \phi_j - \phi_i$ between adjacent qubits to specific values (e.g., $\Delta\phi_{ij} = \pi/4, \pi/2, ...$), the XY+DM interactions with controllable ratios $G/D = \cot(\Delta\phi_{ij})$ can be engineered, yielding characteristic values (of 1,0,...) for the given phase differences.
We experimentally realize the XY+DM interaction in a one-dimensional spin chain with periodic boundary conditions, arranged in a ring geometry, where the XY+DM interaction can be achieved within a single cycle. As illustrated in Fig. 2(a), we implement local $R_\mathrm{z}$ modulation with equally-spaced phase differences (e.g., $\Delta\phi_{ij} = \pi/4, \pi/2, ... $) using $V_{\mathrm{z}}$ gates, across an 8-qubit system to engineer the target interaction.

To validate the engineered \(G/D\) interaction ratio, we prepare the system in the eigenstate $|\psi_0\rangle = \mathcal{V} (G/D)|\psi_x\rangle$, where $|\psi_x\rangle$ represents an $x$-polarized ferromagnetic state. The unitary transformation
$\mathcal{V}(G/D) = \bigotimes_{l=1}^L e^{i l (\phi/2) \sigma_l^z}$, with $\phi = \tan^{-1}(G/D) + \pi$, generates a spin spiral texture characterized by wavevector $\phi$, analogous to classical chiral order~\cite{Garate2010, Kargarian2009}. This zero-energy eigenstate of Eq.~\eqref{Eq_DM_XY} serves as a sensitive fidelity probe: its dynamical stationarity directly reflects the accuracy of the engineered Hamiltonian.  

In Figs.~\ref{fig:DMz}(d) and (e), we display the time-resolved local magnetization \(\langle \tilde{S}^x(t) \rangle = \frac{1}{L} \sum_{i=1}^L \langle \tilde{\psi}(t) | S_i^x | \tilde{\psi}(t) \rangle\), where
\(|\tilde{\psi}(t)\rangle = \mathcal{V}^{-1}(G/D)|\psi(t)\rangle\). For $G/D = 0$ corresponding to pure DM interaction, realized by imposing a $\pi/2$ phase difference between adjacent qubits via virtual Z gates, the spin dynamics (Fig.~\ref{fig:DMz}(d), blue curve) exhibit striking freezing behavior, a hallmark of eigenstate stationarity. In contrast, the evolution of non-eigenstates (yellow curve) exhibits significant temporal suppression. 
The similar results are observed for the $G/D = 1$ case, which is achieved by applying a $\pi/4$ phase difference between neighbour qubits. Here, the eigenstate dynamics (Fig.~\ref{fig:DMz}(e), blue curve) also display suppressed temporal variation.  Experimental data display gradual magnetization decay, primarily attributed to control imperfections and decoherence-induced phase errors. 



In summary, we have experimentally demonstrated precise control of multiple tunable spin interactions, including the continuously adjustable XYZ interaction, transverse-field Ising interaction, and XY+DM interaction. Through optimized global and local microwave modulation schemes, these interactions are realized with high precision preciseness, showing excellent agreement with theoretical simulations. Our work extends the capabilities of  quantum simulation in superconducting circuits, providing access to a broader class of exotic quantum spin models for experimental investigation.

See the supplementary material for detailed information on the theoretical model and device parameters.

We thank H. Zhao for helpful discussions. We thank the support from the Synergetic Extreme Condition User Facility (SECUF) in Huairou District, Beijing. Devices were made at the Nanofabrication Facilities at Institute of Physics, CAS in Beijing. We thank CETC 16 for providing the dilution refrigerator, enabling reliable sample testing through its extreme low-temperature and stable operation.
This work was supported by National Natural Science Foundation of China (Grants Nos.~\!92265207, T2121001, 12122504, T2322030, 92365301, 12404578, 12447184), the Innovation Program for Quantum Science and Technology (Grant No.~\!2021ZD0301800), Beijing Nova Program (Nos.~\!20220484121, 2022000216), Beijing National Laboratory for Condensed Matter Physics (2024BNLCMPKF022), and the China Postdoctoral Science Foundation (Grant No. GZB20240815, GZC20252227).



%

\end{document}


\title{Supplementary Material: "Microwave Engineering of Tunable Spin Interactions with Superconducting Qubits"}

\author{Kui Zhao}
    \thanks{These authors contributed equally to this work.}
	\affiliation{\addrA}

\author{Ziting Wang}
    \thanks{These authors contributed equally to this work.}
	\affiliation{\addrA}
	
\author{Yu Liu}
    \email{yuliu@iphy.ac.cn}
    \thanks{These authors contributed equally to this work.}
	\affiliation{\addrB}
	\affiliation{\addrC}
	
\author{Gui-Han Liang}
    \thanks{These authors contributed equally to this work.}
	\affiliation{\addrB}
	\affiliation{\addrC}
	
\author{Cai-Ping Fang}
	\affiliation{\addrB}
	\affiliation{\addrC}
	
\author{Yun-Hao Shi}
	\affiliation{\addrB}
	
\author{Lv Zhang}
	\affiliation{\addrB}
	\affiliation{\addrC}   
    
\author{Jia-Chi Zhang}
	\affiliation{\addrB}
	\affiliation{\addrC}
    
\author{Tian-Ming Li}
	\affiliation{\addrB}
	\affiliation{\addrC}	

\author{Hao Li}
	\affiliation{\addrA}

\author{Yueshan Xu}
	\affiliation{\addrA}
	
\author{Wei-Guo Ma}
	\affiliation{\addrB}
	\affiliation{\addrC}

\author{Hao-Tian Liu}
	\affiliation{\addrB}
	\affiliation{\addrC}

\author{Jia-Cheng Song}
	\affiliation{\addrB}
	\affiliation{\addrC}

\author{Zhen-Ting Bao}
	\affiliation{\addrB}
	\affiliation{\addrC}

\author{Yong-Xi Xiao}
	\affiliation{\addrB}
	\affiliation{\addrC}

\author{Bing-Jie Chen}
	\affiliation{\addrB}
	\affiliation{\addrC}

\author{Cheng-Lin Deng}
	\affiliation{\addrB}
	\affiliation{\addrC}
	
\author{Zheng-He Liu}
	\affiliation{\addrB}
	\affiliation{\addrC}	

\author{Yang He}
	\affiliation{\addrB}
	\affiliation{\addrC}
	
\author{Si-Yun Zhou}
	\affiliation{\addrB}
	\affiliation{\addrC}	
	
\author{Xiaohui Song}
	\affiliation{\addrB}
	\affiliation{\addrC}

\author{Zhongcheng Xiang}
	\affiliation{\addrB}
	\affiliation{\addrC}
	\affiliation{\addrD}

\author{Dongning Zheng}
	\affiliation{\addrB}
	\affiliation{\addrC}
	\affiliation{\addrD}
	\affiliation{\addrE}
	
\author{Kaixuan Huang}
    \email{huangkx@baqis.ac.cn}
	\affiliation{\addrA}
	
\author{Kai Xu}
    \affiliation{\addrB}
	\affiliation{\addrA}
	\affiliation{\addrC}
	\affiliation{\addrD}
	\affiliation{\addrE}
    
\author{Heng Fan}
    \email{hfan@iphy.ac.cn}
	\affiliation{\addrB}
	\affiliation{\addrA}
	\affiliation{\addrC}
	\affiliation{\addrD}
	\affiliation{\addrE}

\maketitle


\section{Concise Protocol for Mapping Time Intervals to XYZ Model Coefficients\label{app:XYZ}}

Here, we establish a concise protocol to map the time intervals $\tau_i$ ($i=1,2,3$) to the coefficients $J_i$ ($i=x,y,z$) of the XYZ model. Through time-averaging over a periodically driven sequence, a time-independent effective Hamiltonian follows~\cite{science.abd9547,PRXQuantum.3.020303}:
\begin{equation}\label{eq:Hav_001}
\begin{aligned}
H_{\mathrm{av}}&=\frac{1}{T_c} \sum_{\langle ij \rangle} J_{i j} \left[ \tau_1 \left( 
 \sigma^{x}_{i} \sigma^{x}_{j}+\sigma^{y}_{i} \sigma^{y}_{j}\right) + \tau_2 \left(\sigma^{x}_{i} \sigma^{x}_{j}+\sigma^{z}_{i} \sigma^{z}_{j} \right) \right. \\ 
&\quad\quad\quad \left. +\tau_3 \left(\sigma^{y}_{i} \sigma^{y}_{j}+\sigma^{z}_{i} \sigma^{z}_{j} \right) \right]
\end{aligned}
\end{equation}
where $T_c=2\left(\tau_1+\tau_2+\tau_3\right)$ is the total duration of a sequence. The system dynamics are well-approximated by $H_{\mathrm{av}}$. To establish the correspondence between the time intervals $\tau_i(i=1,2,3)$ and the XYZ model coefficients $\delta_i(i=x,y,z)$, an invertible transformation matrix $\mathcal{T}$ is introduced,
\begin{equation}\label{eq:Hav_002}
\mathcal{T}\left(\begin{array}{l}
\sigma^{x}_{i} \sigma^{x}_{j}+\sigma^{y}_{i} \sigma^{y}_{j} \\
\sigma^{x}_{i} \sigma^{x}_{j}+\sigma^{z}_{i} \sigma^{z}_{j} \\
\sigma^{y}_{i} \sigma^{y}_{j}+\sigma^{z}_{i} \sigma^{z}_{j}
\end{array}\right)=\left(\begin{array}{c}
\sigma^{x}_{i} \sigma^{x}_{j} \\
\sigma^{y}_{i} \sigma^{y}_{j} \\
\sigma^{z}_{i} \sigma^{z}_{j}
\end{array}\right)
\end{equation}
with
\begin{equation}\label{eq:Hav_003}
\mathcal{T}=1/2\left(\begin{array}{ccc}
1 & 1 & -1 \\
1 & -1 & 1 \\
-1 & 1 & 1
\end{array}\right)
\end{equation}

Rewriting $H_{\mathrm{av}}$ in matrix form,
\begin{equation}\label{eq:Hav_004}
\begin{aligned}
H_{\mathrm{av}} &= {J_{ij}} \left(\frac{\tau_1}{T_c}, \frac{\tau_2}{T_c}, \frac{\tau_3}{T_c}\right) \left(\begin{array}{l}
\sigma^{x}_{i} \sigma^{x}_{j}+\sigma^{y}_{i} \sigma^{y}_{j} \\
\sigma^{x}_{i} \sigma^{x}_{j}+\sigma^{z}_{i} \sigma^{z}_{j} \\
\sigma^{y}_{i} \sigma^{y}_{j}+\sigma^{z}_{i} \sigma^{z}_{j}
\end{array}\right) \\
&= {J_{ij}} \left(\frac{\tau_1}{T_c}, \frac{\tau_2}{T_c}, \frac{\tau_3}{T_c}\right) T^{-1} \left(\begin{array}{c}
\sigma^{x}_{i} \sigma^{x}_{j} \\
\sigma^{y}_{i} \sigma^{y}_{j} \\
\sigma^{z}_{i} \sigma^{z}_{j}
\end{array}\right) \\
&\equiv {J_{ij}} \left(\delta_x, \delta_y, \delta_z\right)  \left(\begin{array}{c}
\sigma^{x}_{i} \sigma^{x}_{j} \\
\sigma^{y}_{i} \sigma^{y}_{j} \\
\sigma^{z}_{i} \sigma^{z}_{j}
\end{array}\right)
\end{aligned}
\end{equation}
we recover the XYZ Hamiltonian,
\begin{equation}\label{eq:Hav_005}
H_{\mathrm{XYZ}}= \sum_{\langle ij \rangle} J_{i j}  \left(\delta_x \sigma^{x}_{i}\sigma^{x}_{j} +\delta_y \sigma^{y}_{i} \sigma^{y}_{j}+\delta_z \sigma^{z}_{i} \sigma^{z}_{j}\right)
\end{equation}
where the coefficients satisfy,
\begin{equation}\label{eq:Hav_006}
\left(\frac{\tau_1}{T_c}, \frac{\tau_2}{T_c}, \frac{\tau_3}{T_c}\right) \mathcal{T}^{-1} \equiv \left(\delta_x, \delta_y, \delta_z\right).
\end{equation}

Given the invertibility of the $\mathcal{T}$ matrix, the relationship between $\tau_i$ and $\delta_i$ is inherently direct. Specifically, the $\mathcal{T}$ matrix provides a systematic mapping from the weighted combinations of spin-spin interaction terms in the averaged Hamiltonian $H_{\mathrm{av}}$ to the individual anisotropy parameters $\delta_i$. This formalism enables precise engineering of XYZ coefficients through tailored time sequences. Moreover, Eq.~(\ref{eq:Hav_006}) explicitly imposes the constraint $\delta_x + \delta_y + \delta_z = 1$, which fundamentally limits the range of attainable anisotropy configurations.

\begin{table}
\caption{\label{tab:table1}
	\small{\textbf{Basic parameters of eight qubits Q$_1$-Q$_8$ used in the present experiment.} $\omega_{r}$, $\omega_{\rm{max}}$, and $\omega_{\rm{idle}}$ are the readout resonator frequency, the qubit maximum frequency, and the qubit idle frequency, respectively. $\alpha$ is the qubit anharmonicity. $T_{1}$ and $T_2^{\ast}$ are the energy relaxation and dephasing time of the qubit at idle point. $F_{g}$ and $F_{e}$ are the readout fidelities for the ground and first-excited states. The infidelities of the single-qubit gate are calibrated by RB. }}
    \begin{tabular}{c|cccccccc}
    		\hline\hline
    		Qubits & $\rm Q_1$ & $\rm Q_2$ & $\rm Q_3$ & $\rm Q_4$ & $\rm Q_5$ & $\rm Q_6$& $\rm Q_7$& $\rm Q_8$\\ 
      \hline
      $\omega_r$ (GHz)& 7.346	&7.352	&7.356	&7.315&	7.302&	7.245&	7.285	&7.332 \\
      $\omega_{\rm max}$ (GHz) & 5.058&	5.234&	5.119&	5.149&	5.076&	5.071&	4.986&	5.065\\
      $\omega_{\rm idle}$ (GHz) &5.058 & 4.484	&5.106 &	4.381&	5.038&	4.572&	4.986&	4.434\\
      $\alpha_{\rm idle}$ (MHz)&-192.8&	-211.1&	-196.6&	-210.8&	-197.7&	-210.1&	-197.3&	-209.9\\
      
      $T_{1}$ ($\mu$s)&22.26&	30.31&	18.80&	46.34&	24.24&	21.79&	21.24&	60.76\\
      
      $T^*_{2}$($\mu$s)&10.87&	0.84&	6.48&	0.97&	2.53&	1.17&	14.92&	1.24\\
      
      $F_g$($\%$)&96.6 &94.4&	94.0&	97.5&	96.4&	96.7&	90.9&	95.4	\\
      $F_e$($\%$) & 94.1 &91.1&	90.1&	94.8&	95.0&	94.9&	84.9&	93.1\\
      
      $\omega_{\rm work}$ (GHz)& 4.65& 4.65& 4.65& 4.65& 4.65& 4.65& 4.65& 4.65\\
     1Q RB ($\%$)& 99.971& 99.849& 99.943& 99.906& 99.959& 99.886& 99.941& 99.889\\
  \hline\hline
	\end{tabular}
\end{table}

\section{Device parameters}
The experiment utilized a quantum device based on a two-dimensional 9-qubit chip, where eight peripheral qubits (Q$_1$-Q$_8$) were selected to form a ring configuration. As shown in Table~.\ref{tab:table1}, key parameters of the qubits: The maximum frequencies ($\omega_{\rm{max}}$) of the qubits range from 4.986 GHz to 5.234 GHz, while their idle frequencies ($\omega_{\rm{idle}}$) span 4.381–5.106 GHz. All qubits were operated at a uniform working frequency of $\omega_{\rm{work}} = 4.65$ GHz. For decoherence times, the energy relaxation time $T_1$ ranged from 18.80 $\mu$s to 60.76 $\mu$s, with Q$_8$ exhibiting the longest $T_1 = 60.76$ $\mu$s. The dephasing time $T_2^*$ varied significantly (0.84–14.92 $\mu$s). The readout fidelities for the ground state ($F_g$) and first-excited state ($F_e$) both achieved around 90$\%$. Single-qubit gates, calibrated via randomized benchmarking (RB), demonstrated an average fidelity of approximately 99.9$\%$.



\section{Experimental Hamiltonian and Rotating Frame Transformations}

In this section, we present an analysis of the physical mechanism underlying the implementation of frame phase compensation for the XYZ model.
The experimental Hamiltonian of the qubit system in the laboratory frame is defined piecewise over consecutive time intervals as follows:  

\begin{equation}
H = 
\begin{cases} 
H_0^{\text{idle}} + H_d(t), & \text{for time interval } t_1, \\
H_0^{\text{work}} + H_{\text{XY}}, & \text{for time interval } t_2, \\
H_0^{\text{idle}} + H_d(t), & \text{for time interval } t_3, \\
\vdots
\end{cases}
\label{C1}
\end{equation}  
where, the static Hamiltonians,  
\begin{equation}
H_0^{\text{idle}} = \sum_{i} \frac{\omega_i^{\text{idle}}}{2} \sigma_i^z \quad \text{and} \quad H_0^{\text{work}} = \sum_{i} \frac{\omega_i^{\text{work}}}{2} \sigma_i^z,
\label{C2}
\end{equation}  
the drive Hamiltonian,
\begin{equation}
H_d(t) = \sum_{i} \Omega_i \left[ \cos(\omega_i^{\text{idle}} t + \phi_i) \frac{\sigma_i^x}{2} + \sin(\omega_i^{\text{idle}} t + \phi_i) \frac{\sigma_i^y}{2} \right],
\label{C3}
\end{equation}  
and the interaction Hamiltonian,  
\begin{equation}
H_{\mathrm{XY}} =  \sum_{\langle ij \rangle} \frac{J_{i,j}}{2} \left( \sigma_i^x\sigma_j^x + \sigma_i^y\sigma_j^y \right).
\label{C5}
\end{equation}

To analyze the system dynamics, we employ rotating frames tailored to the idle and work frequencies. For the  idle frame, applying the unitary transformation\cite{PhysRevA.105.012418} \(|\widetilde{\psi}(t)\rangle^{\text{idle}} = e^{i H_0^{\text{idle}} t} |\psi(t)\rangle\) (with \(H = H_0^{\text{idle}} + H_d(t)\)) eliminates the static term ($\widetilde{H}^{\text{idle}} = e^{i H_0^{\text{idle}} t} H e^{-i H_0^{\text{idle}} t} - i e^{i H_0^{\text{idle}}t} \frac{\partial}{\partial t} \left( e^{-i H_0^{\text{idle}} t} \right)$), reducing the Hamiltonian to:  
\begin{equation}
\widetilde{H}^{\text{idle}} = \sum_{i} \frac{\Omega_i}{2} \left( \cos\phi_i \cdot \sigma_i^x + \sin\phi_i \cdot \sigma_i^y \right).
\end{equation}  
For the work frame, the transformation \(|\widetilde{\psi}(t)\rangle^{\text{work}} = e^{i H_0^{\text{work}} t} |\psi(t)\rangle\), yielding \(\widetilde{H}_{\text{XY}}^{\text{work}} = H_{\text{XY}}\). For coherent analysis of the Floquet dynamics, we anchor the description in the work frame. Transforming the idle interval dynamics to work frame yields the state evolution,
\begin{equation}
\begin{aligned}
|\widetilde{\psi}(t_1)\rangle^{\text{work}} &= e^{iH_0^{\text{work}} t_1} |\psi(t_1)\rangle \\
&= e^{i(H_0^{\text{work}} - H_0^{\text{idle}}) t_1} |\widetilde{\psi}(t_1)\rangle^{\text{idle}} \\
&= e^{i(H_0^{\text{work}} - H_0^{\text{idle}}) t_1} e^{-i\widetilde{H}^{\text{idle}} t_1} |\psi(0)\rangle,
\end{aligned}
\end{equation}  
where \(e^{-i \widetilde{H}^{\text{idle}} t_1}\) represents single-qubit rotations (e.g., \(\pi/2\) pulses), and \(e^{i (H_0^{\text{work}} - H_0^{\text{idle}}) t_1}\) accounts for phase accumulation from frame transitions (due to frequency shifts). In XYZ model implementations, this phase must be an integer multiple of \(2\pi\) for per qubit. 
In experimental implementations, physical Z gates (e.g., via $A_{\mathrm{zpulse}}$-tuned $R_z$ gates) can be employed to compensate for this phase difference during runtime, while virtual Z gates can be effectively employed in the initial stage to compensate for the phase accumulation by predefining the coordinate system.

For a Floquet period $T_c$, the unitary evolution incorporating repeated idle-work transitions involves:  
\begin{equation}
\begin{aligned}
U(T_c) &= \dotsm \underbrace{e^{-i (H_0^{\text{work}} - H_0^{\text{idle}}) t_3}}_{\text{Real-time} \;R_z\; \text{compensation}} e^{i \widetilde{H}^{\text{idle}} t_3}   e^{-i H_{\text{XY}} t_2}  \underbrace{e^{-i (H_0^{\text{work}} - H_0^{\text{idle}}) t_1}}_{\text{Real-time} \;R_z\; \text{compensation}} e^{i \widetilde{H}^{\text{idle}} t_1}\\
&= \dotsm {e^{i \widetilde{H}^{\text{idle}} t_3} e^{-i H_{\text{XY}} t_2} e^{-i \widetilde{H}^{\text{idle}} t_1}}.
\end{aligned}
\end{equation}  
When the idle drives implement \(\pi/2\) rotations (e.g., $e^{i\widetilde{H}^{\text{idle}} t_1} =  e^{\frac{\pi}{2} \cdot \sum_i \sigma_i^x /2}$), the evolution reshapes the XY Hamiltonian via collective  global rotations:  
\begin{equation}
U(T) = \dotsm e^{i \frac{\pi}{2} (\sigma_i^x/2 + \sigma_j^x/2)} e^{-i H_{\text{XY}} t_2} e^{-i \frac{\pi}{2} (\sigma_i^x/2 + \sigma_j^x/2)} = \do R^{\mathbf{x}}_{i,j}(\theta)(\pi/2) e^{-i H_{\text{XY}} t_2} R^{\mathbf{x}}_{i,j}(\theta)(-\pi/2),
\end{equation}   
where $R^{\mathbf{x}}_{i,j}(\theta) = \exp\left(-i \frac{\theta}{2} (\sigma_i^x + \sigma_j^x)\right)$. This operation effectively transforms the XY interaction into alternative forms (e.g., XZ or YZ), enabling the design of tailored effective Hamiltonians in the Floquet framework through precise phase compensation and drive engineering.

\section{System Size Constraints for Periodic Boundary Conditions}  
\label{supp:size_constraints}  

In our experiment, we selected an 8-qubit subset arranged in a ring geometry to realize periodic boundary conditions for the XY+DM interaction. This configuration requires the total phase accumulation around the ring to satisfy $\sum \Delta\phi_{ij} = 2\pi n$ ($n \in \mathbb{Z}$), which imposes specific constraints on system size.
The Hamiltonian $\hat{H}_{XY+DM}$ is engineered through local modulations:
\begin{equation}
\hat{H}_{XY+DM} = \sum_{\langle ij \rangle} R_{i,j}^z(\phi_i,\phi_j) \hat{H}_{i,j}^{XY} R_{i,j}^z(-\phi_i,-\phi_j)
\end{equation} 
where the phase difference $\Delta\phi_{ij} \equiv \phi_j - \phi_i$ determines the $G/D$ ratio via $G/D = \cot(\Delta\phi_{ij})$. For periodic boundary conditions, the system size $N$ must satisfy $N \cdot \Delta\phi_{ij} = 2\pi n$. 

Specific implementations require:

- For $G/D=1$ ($\Delta\phi_{ij} = \pi/4$): $N$ must be a multiple of 8

- For $G/D=0$ ($\Delta\phi_{ij} = \pi/2$): $N$ must be a multiple of 4

The 8-qubit ring satisfies both conditions (8 is divisible by both 4 and 8), enabling implementation of both interaction regimes within the same geometry.

\color{black}

%